%% file: ICRC2023_template_IceCube.tex
\documentclass[a4paper,11pt]{article}
\usepackage{siunitx}
\usepackage{wrapfig}
\usepackage{float}
\usepackage{subfig}
\usepackage{pos}
 \usepackage{chemformula}
\usepackage{url}
\usepackage{lipsum} %package to generate placeholder text in the following
\usepackage{lineno}
\title{Detailed investigations of PMTs in optical sensors for neutrino telescopes such as IceCube Upgrade}

\ShortTitle{Detailed investigations of PMTs}

\author{The IceCube Collaboration \\{\normalsize \normalfont(a complete list of authors can be found at the end of the proceedings)}\\}

\emailAdd{berit.schlueter@icecube.wisc.edu}
\emailAdd{w\_acht02@uni-muenster.de}
\emailAdd{m.unland@uni-muenster.de}

\abstract{

Photomultiplier tubes (PMTs) are a central component of neutrino telescopes such as IceCube and KM3NeT, and an accurate understanding and measurement of their properties is indispensable for improvements of these experiments. In this contribution we focus on a detailed investigation of the photocathode and the dynode system and their influence on the performance of the PMT. Three methods are used for the investigation. Ellipsometry measurements of the photocathode analyze its optical properties in terms of absorption probability and refractive index. Scans of the photocathode in single photon illumination probe performance differences along the photocathode surface. Systematic deviations in the resulting amplifications are compared to electric field and electron tracing simulations through the dynode system to understand the measured values. 
The goal is an extensive understanding of efficiency, amplification, and timing as functions of wavelength and impact point as well as angle.
\vspace{4mm}
{\bfseries Corresponding authors:}
Berit Schlüter$^{1*}$, Willem Achtermann$^{1}$, Martin Antonio Unland Elorrieta$^{1}$\\
{$^{1}$ \itshape Institut für Kernphysik, Westfälische Wilhelms-Universität Münster, Münster, Germany}\\[4mm]
$^*$ Presenter

\ConferenceLogo{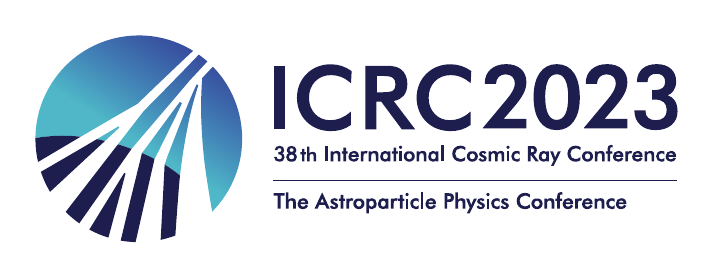}

\FullConference{The 38th International Cosmic Ray Conference (ICRC2023)\\ 26 July -- 3 August, 2023\\ Nagoya, Japan}
}

\begin{document}

\maketitle
%%%%%%%%%%%%%%%%%%%%%%%%%%%%%%%%%%%%%%%%%%%%%%%%%%%%%%%%%%%%%%%
\section{Introduction}\label{sec:Introduction}
Current neutrino telescopes consist of an array of digital optical modules containing one or more photomultiplier tubes (PMTs). One of these experiments is the IceCube Neutrino Observatory~\cite{Aartsen:2016nxy} located at the South Pole. IceCube is a Cherenkov detector consisting of 86 strings with 5160 digital optical modules in total that can detect neutrinos in the energy range from about $\SI{\sim10}{\giga\electronvolt}$ to several $\si{\peta\electronvolt}$ and reconstruct their direction and energy from the amount and arrival times of the photons measured by the DOMs. Therefore, an accurate understanding of PMT performance is key for the correct reconstruction of events. An extension of seven additional strings is planned, called the IceCube Upgrade, to reduce the energy threshold of the detector to about $\SI{\sim1}{\giga\electronvolt}$ and to more accurately calibrate the current detector~\cite{IceCubeUpgrade}. The IceCube Upgrade will include new types of optical modules, one of them is the multi-PMT digital optical module (mDOM). It consists of 24 PMTs of $\SI{80}{\milli\meter}$ diameter each, achieving a more than a factor two larger total photocathode area compared to the current IceCube DOM while also providing intrinsic angular resolution. 

PMTs are used to detect the light by converting it into an electrical signal. A PMT is a vacuum tube containing a photocathode and a multiplication system consisting of a series of dynodes. The vacuum housing is commonly made of borosilicate glass. The photocathode is a semi-transparent thin layer of a compound semiconductor evaporated on the inside of the glass window of the PMT. The mDOM PMT uses biakali as the photocathode material. When a photon hits the PMT, it can be converted into a primary electron in the photocathode by the photoelectric effect. The primary electron is then accelerated to the dynode system where it is multiplied. At the end, the secondary electrons are collected at the anode.\\
\\
In this work, three methods for the detailed study of the performance of the mDOM PMTs (Hamamatsu R15458-02 \cite{MartinDoctorialThesis}) are presented, namely a PMT inhomogeneity study (see chapter \ref{sec:PMT}), the photocathode optical properties study (see chapter \ref{sec:ellipsometry}), and the electron tracing simulation (see chapter \ref{sec:comsol}). 
%%%%%%%%%%%%%%%%%%%%%%%%%%%%%%%%%%%%%%%%%%%%%%%%%%%%%%%%%%%%%%%
\section{PMT inhomogeneity studies}\label{sec:PMT}
 
\begin{figure}[tb]
\centering
\subfloat[\label{subfig:PMT_orientation}]{%
     \includegraphics[width=0.41
     \textwidth]{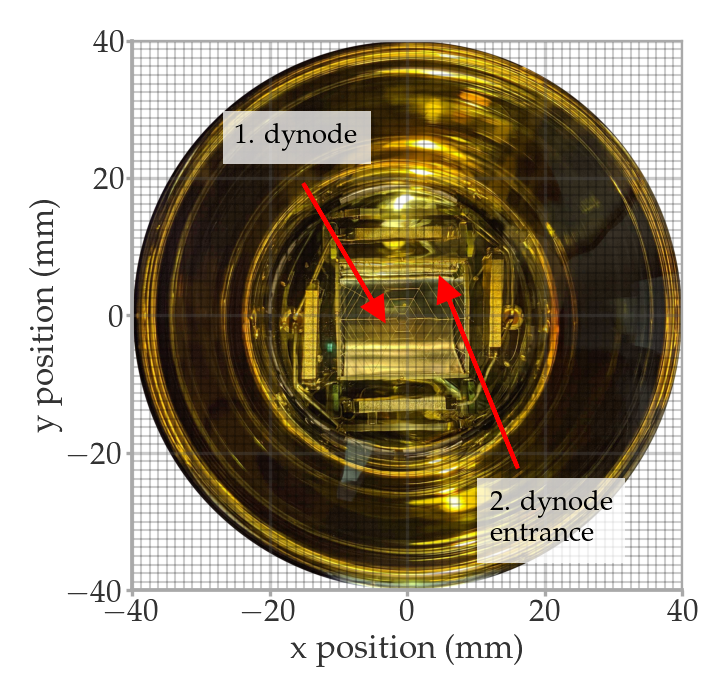}
     }
\subfloat[\label{subfig:rel_TT_scan}]{%
       \includegraphics[width=0.49\textwidth]{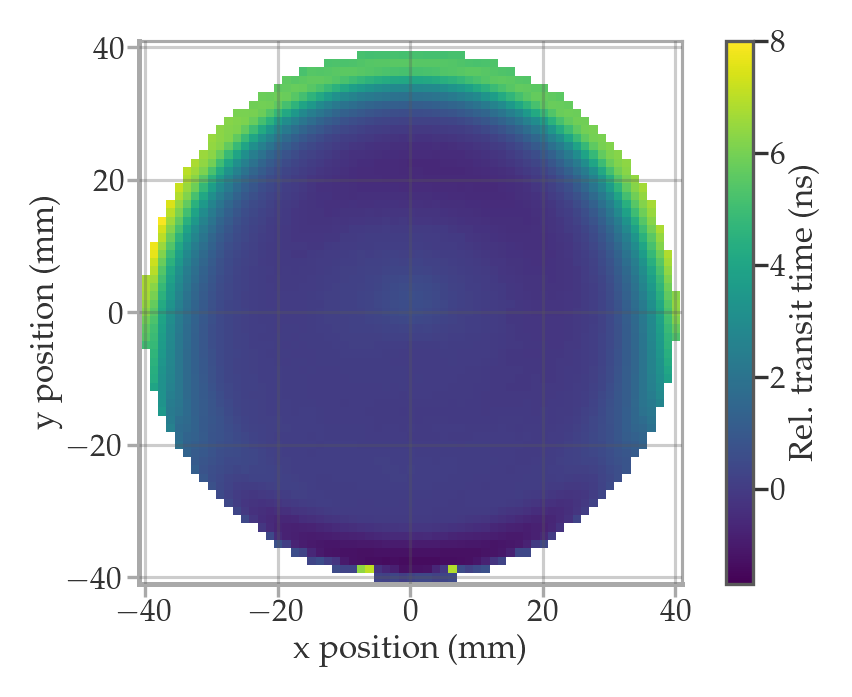}
     }
  \caption{a): Orientation of the internal components of the PMT during the scans. b): Relative transit time scan of a PMT at $\SI{459}{\nano\meter}$. Figure taken from \cite{MartinDoctorialThesis}.}
  \label{fig:PMT_scan}
\end{figure}

When characterizing the properties of PMTs, only the average parameters are usually measured, since the entire photocathode surface is illuminated during the measurements. To study the inhomogeneities of a PMT, the photocathode of the PMT can be scanned with collimated light. In \cite{MartinDoctorialThesis,MartinScan} the PMT was scanned in a grid of $\SI{1.2}{\milli \meter}$ $\times$ $\SI{1.2}{\milli\meter}$ to measure several performance parameters. The PMT always had the internal component orientation shown in figure \ref{subfig:PMT_orientation}. As an example, the results for the relative transit time are shown in figure \ref{subfig:rel_TT_scan}. The transit time is the time difference between absorption of the photon at the photocathode and time of the PMT signal. In this case the transit time is measured at a wavelength of $\SI{459}{\nano\meter}$ and plotted relative to the mean of the central area of the PMT (radius < $\SI{30}{\milli\meter}$).\begin{figure}[tb]
\captionsetup[subfloat]{labelformat=empty}
     \subfloat[\label{subfig:eff_460nm}]{%
     \includegraphics[trim={0.3cm 0.24cm 0.3cm 0.24cm},clip,width=0.325\textwidth]{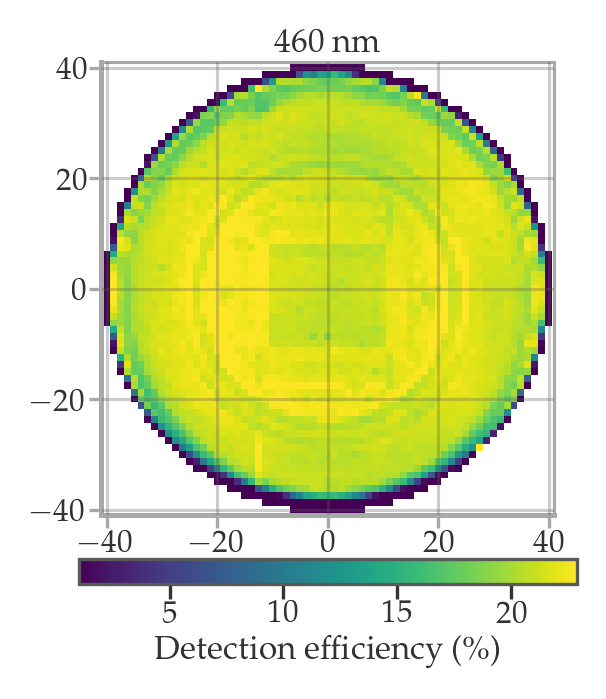}
     }
      \subfloat[\label{subfig:eff_540nm}]{%
       \includegraphics[trim={0.3cm 0.24cm 0.3cm 0.24cm},clip,width=0.325\textwidth]{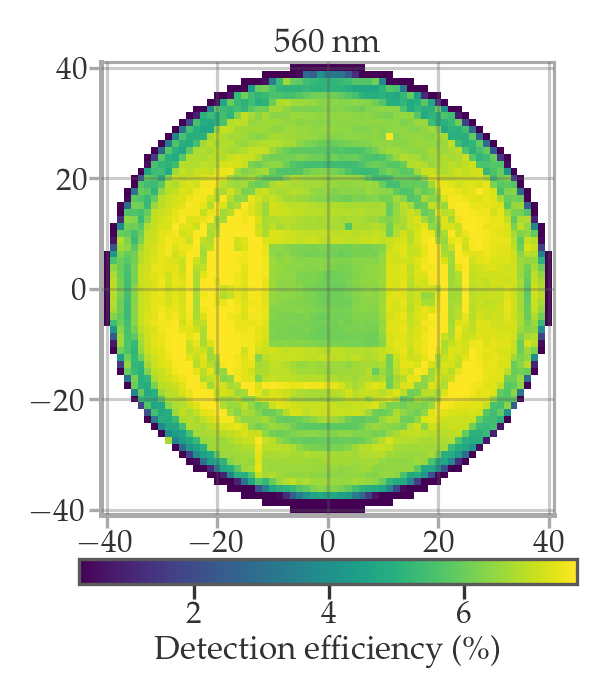}
     }
     \subfloat[\label{subfig:eff_620nm}]{%
       \includegraphics[trim={0.3cm 0.24cm 0.3cm 0.24cm},clip,width=0.325\textwidth]{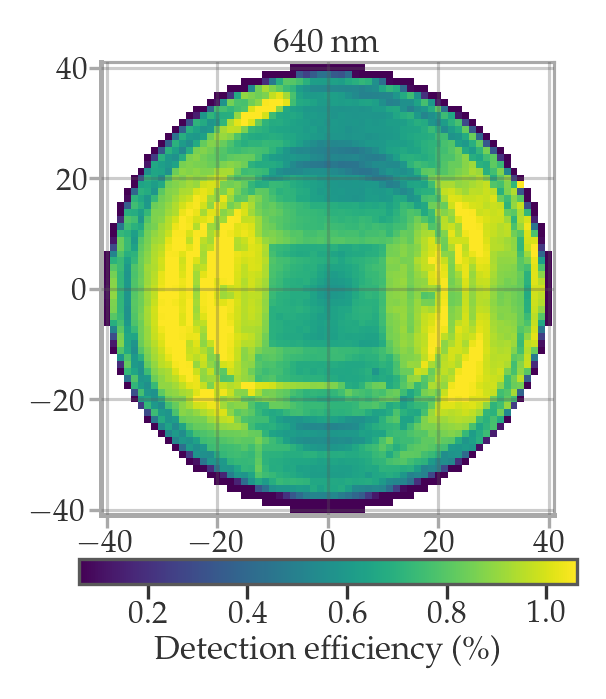}
     }
     \vspace{-0.5cm}
     \caption{Detection efficiency scan of an mDOM PMT for $\SI{460}{\nano\meter}$, $\SI{560}{\nano\meter}$ and $\SI{640}{\nano\meter}$. Figure taken from \cite{MartinDoctorialThesis}. }
     \label{fig:wavelength_scans}
   \end{figure} It can be observed, that the transit time increases at the edge of the PMT, except of the case of the negative y-values. This is believed to be caused by the different paths that the primary electrons can take on the way to the first dynode.
 \\
 \\
The homogeneity studies were also performed at different wavelengths in \cite{MartinDoctorialThesis}. Figure \ref{fig:wavelength_scans} shows the detection efficiency at $\SI{460}{\nano\meter}$, $\SI{540}{\nano\meter}$ and $\SI{620}{\nano\meter}$. For $\SI{460}{\nano\meter}$, it can be seen that the detection efficiency decreases at the edge of the PMT. The internal components of the PMT are also visible due to light reflection from these components. It can be noticed that the homogeneity of the detection efficiency decreases with increasing wavelength. Additional structures can also be seen next to the structures of the internal components for higher wavelengths. These new structures are thought to be due to differences in the thickness of the photocathode along the PMT \cite{MartinDoctorialThesis}.  \\
 \\
The results of the scans motivate to investigate the PMTs in more detail. On the one hand with respect to the photocathode and its optical properties and thickness and their influence on the PMT efficiency. On the other hand, with respect to the dependence of pulse parameters on the trajectories of the primary electron inside the PMT. The first results of these two new studies will be described in the chapters below.
%%%%%%%%%%%%%%%%%%%%%%%%%%%%%%%%%%%%%%%%%%%%%%%%%%%%%%%%%%%%%%%
\section{Measurement of the optical properties of the photocathode}\label{sec:ellipsometry}

One way to investigate the PMT in more detail is to analyze the optical properties of the photocathode, such as the complex refractive index $n^*$ and thickness $d$, to gain a better understanding of their influence on the PMT efficiency. The quantum efficiency (QE) of a PMT, which describes the ratio of the number of primary electrons to the number of incident photons, can be described by a three step process: the absorption probability of the photocathode, the electron diffusion in the photocathode and the escape probability into the vacuum. The absorption probability can be calculated from the imaginary refractive index and the photocathode thickness and, therefore, they are qualitatively related to QE. Inspired by \cite{Motta2005}, an ellipsometer was commissioned and characterized to measure the optical properties of the photocathode. Using an ellipsometer one is able to analyze the optical properties of thin layers by measuring the change in polarization of light after reflecting at a sample. The light from the laser is passed through a combination of a polarizer and a quarter wave plate to obtain a specific polarization and is analyzed by a second polarizer (called analyzer) after the reflection. The intensity after the analyzer as a function of the polarizer and analyzer angles is measured using a photodiode. Following the methodology of Null-Ellipsometry \cite{Ellipsometry}, the angles resulting in extinction are used to calculate the \textit{ellipsometry parameters} $\psi$ and $\Delta$. These relate to the reflection coefficients $r_{s,p}$ of s- (perpendicular) and p- (parallel) polarized light via 
\begin{align}
    \tan(\psi) = \frac{|r_\text{p}|}{|r_\text{s}|} \ \ \text{and} \ \ \Delta = \delta_{r_\text{p}} - \delta_{r_\text{s}},
    \label{eq:psi_delta}
\end{align}
where $|r_ {\text{p,s}}|$ describes the amplitude and $\delta_{r_\text{p,s}}$ the phase. For any system and incidence angle the expected $r_{s,p}$ for an assumed set of refractive indices and thicknesses can be calculated with Fresnel equations. The true optical properties of the system are deduced by repeating the above measurements and calculations for a large set of incidence angles (see figure \ref{fig:elli_para_plots}). The PMT can be thought as a four layer system and the reflection coefficients are calculated with Fresnel equations. The first layer describes the ambiance with a refractive index of $n_1$ in which the PMT is located. The second layer represents the glass window of the PMT with a refractive index of $n_2^* = n_2 + i \cdot k_2$. The photocathode can be considered as the third layer and has a complex refractive index of $n_3^* = n_3 + i \cdot k_3$ and a thickness of $d$. Since the photocathode is a thin layer (thickness of this layer is much smaller than the wavelength of the laser), the general Fresnel equations must be adapted, since interference can occur within the photocathode. The last layer is the vacuum inside the PMT with a refractive index of $n_4 = 1$. Using the description of the PMT as a four layer system, the reflection coefficient $r_{\text{234}}$ for the reflection at the photocathode can be calculated as follows: 
\begin{align}
    r_\text{234} = \frac{r_{23}+r_{34}\exp(2i\rho)}{1+r_{23}r_{34}\exp(2i\rho)}  \ \ \ \ \
    \label{eq:r234}
 \text{with} \ \ \ \ \
 \rho = \frac{2\pi d}{\lambda} n_3^* \cos(\theta_3),
\end{align}
where $r_{ij}$ is the reflection coefficient at the boundary between layer $i$ and $j$ and $\rho$ describes the phase difference. The phase shift that occurs depends on the wavelength $\lambda$ as well as the angle of incidence on the photocathode $\theta_3$, which can be calculated with the help of Snell's law from the incident angle at the first layer $\theta_1$. For a more in depths understanding, the reader is referred to \cite{Ellipsometry, Motta2005}.
%%%%%%%%%%%%%%%%%%%%%%%%%%%%%%%%%%%%%%%%%%%%%%%%%%%%%%%%%%%%%%%
\newpage
\subsection{Setup}
%\vspace{-4mm}
\begin{wrapfigure}{tb}{0.35\textwidth}
%\vspace{-5mm}
    \centering
    \includegraphics[trim={0.cm 0.cm 1.cm 1.cm},width=0.35\textwidth]{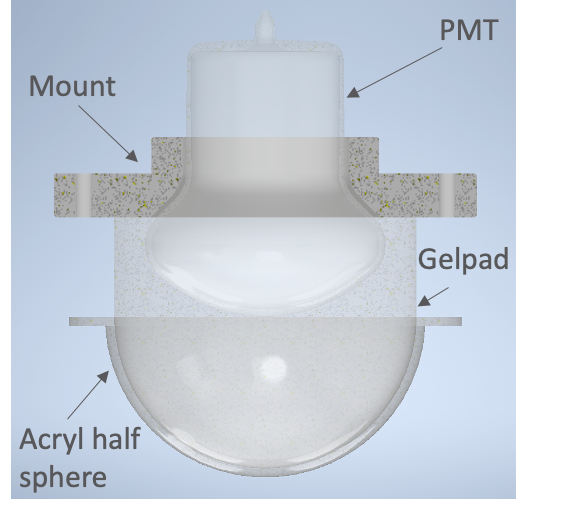}
   \vspace{-9mm}
    \caption{Illustration of the PMT with the gelpad sphere.}
    \label{fig:gelpad}
\end{wrapfigure}

\begin{figure}[!b]
    \centering
     \includegraphics[width=0.80\textwidth]{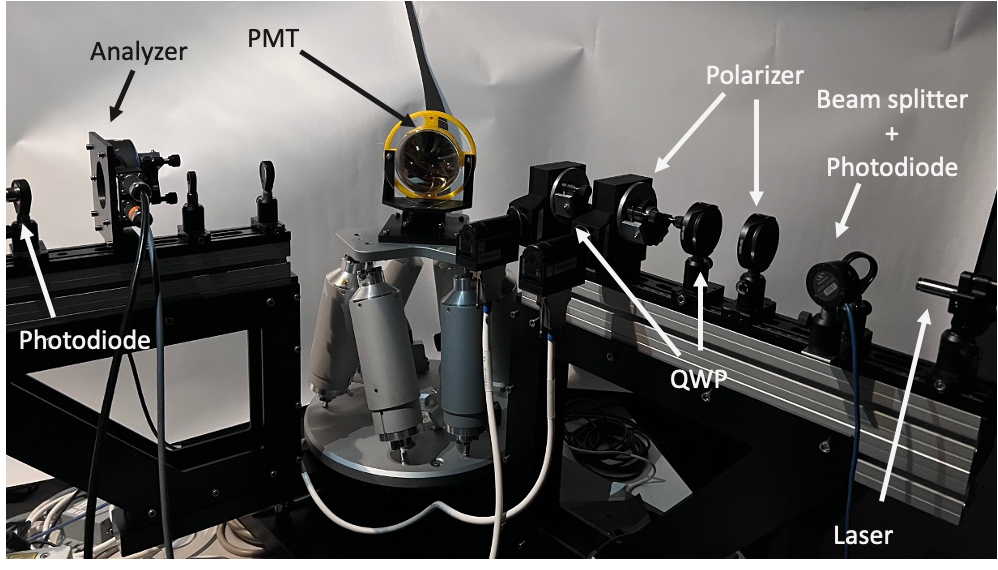}
     \caption{Picture of the used ellipsometry setup. QWP stands for quarter wave plate.}
     \label{fig:setup}
\end{figure}

The sample used is an mDOM PMT with a gel hemisphere (see figure \ref{fig:gelpad}) attached on the center of the photocathode window, which is inserted in the setup shown in figure \ref{fig:setup}. This ensures that the laser beam is always perpendicular to the gel sphere when measuring the center of the photocathode. In addition, the gel sphere has two other advantages: on the one hand, the PMT is in a medium with a refractive index greater than 1, which leads to a peak at the critical angle in the reflection that can be fitted more precisely. On the other hand, the reflection from the glass disappears, so the beam is not displaced by the glass and hits the center of the photocathode directly, as the refractive index of the glass is nearly equal to the refractive index of the gel. The measurements are done at a wavelength of $\SI{480}{\nano \meter}$ for different angles of incidence between $\SI{25}{\degree}$ and $\SI{58}{\degree}$ always measuring the same point at the PMT center. At this wavelength the absorption of the glass bulb is negligible and $k_2$ can be assumed to be $\SI{0}{}$. The setup shown in figure \ref{fig:setup} was characterized with different reference samples in \cite{Masterarbeit}.

%%%%%%%%%%%%%%%%%%%%%%%%%%%%%%%%%%%%%%%%%%%%%%%%%%%%%%%%%%%%%%%
\subsection{Results}
\begin{figure}[H]
\vspace{-5mm}
\captionsetup[subfloat]{labelformat=empty}
     \subfloat[\label{subfig:psi_plot}]{%
     \includegraphics[width=0.49\textwidth]{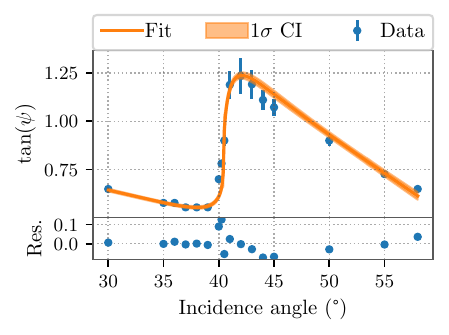}
     }
      \subfloat[\label{subfig:delta_plot}]{%
       \includegraphics[width=0.49\textwidth]{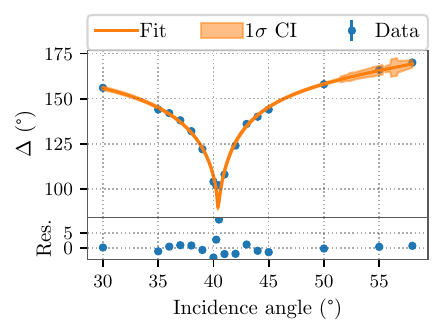}
     }
     \vspace{-0.8cm}
     \caption{Ellipsometry parameters plotted against the incident angle.}
     \label{fig:elli_para_plots}
   \end{figure}

First results of the above described measurement are presented in figure~\ref{fig:elli_para_plots}. The right side shows the tangent of the ellipsometry parameter $\psi$ and the left plot the parameter $\Delta$. These two curves are simultaneously used to fit a number of parameters, among these the complex refractive index $n_3^*$ and the thickness $d$ of the photocathode (see equation \ref{eq:psi_delta} and \ref{eq:r234}). The resulting best estimates are listed in table~\ref{tab:Results_elli} together with reference values from~\cite{Motta2005}. Since the semiconductor mixture of the photocathode is unknown, the reference values of two possible materials at the closest wavelength ($\SI{485}{\nano\meter}$) to the measurement were presented. The thickness of the photocathode is expected to be around $\SI{20}{\nano\meter}$~\cite{Motta2005}.

\begin{table}[h]
    \centering
    \caption{Complex refractive index of two possible photocathode materials and the values of the complex refractive index and thickness of the photocathode from the fit. Values for \ch{K2CsSb} and \ch{RbCsSb} taken from~\cite{Motta2005}.}
    \begin{tabular}{c|c|c|c}
         & refractive index $n$ & extinction coefficient $k$ & thickness $d$  \\
         \hline 
        fit values & $\SI{2.54 \pm 0.08}{}$ & $\SI{1.13 \pm 0.03}{}$ & $\SI{34.8 \pm 1.7}{\nano\meter}$ \\
        \hline
        \ch{K2CsSb} ($\SI{485}{\nano\meter}$)& $\SI{3.00}{}$ & $\SI{1.11}{}$&\\
        \hline
    \ch{RbCsSb} ($\SI{485}{\nano\meter}$)& $\SI{2.99}{}$ & $\SI{1.37}{}$&\\
    \end{tabular}
    
    \label{tab:Results_elli}
\end{table}

The index of refraction is lower than that reported in~\cite{Motta2005}, while the thickness is $\SI{1.5}{}$ times greater than described in~\cite{Wright2017}. These differences could potentially be attributed to variations in photocathode thickness between PMT models or different photocathode compositions. However, these discrepancies might also indicate systematic issues in the setup, such as misalignments or impurities in the gel-pad. To verify that the measurements are reasonable, the absorption of the PMT can be calculated. The absorption must be greater than the QE of the PMT. This is indeed the case, and the absorption agrees with the expectation as calculated from the reference values. Also, refraction measurements for the same incident angle should be incorporated to the analysis to further constrain the fit parameters. In the future, more mDOM PMTs will be measured. The goal is to also determine the thickness variation between different PMTs.
%results
%%%%%%%%%%%%%%%%%%%%%%%%%%%%%%%%%%%%%%%%%%%%%%%%%%%%%%%%%%%%%%%
\section{Electron tracing simulation}\label{sec:comsol}

Another way to study the PMT in more detail is to simulate the electron trajectory inside the PMT. The goal of this simulation is to gain a better understanding of the electron trajectory inside the PMT and the influence of the starting parameters of the primary electron. The simulations are carried out with COMSOL Multiphysics\textsuperscript{\textregistered}, which uses the finite element method (FEM) to solve the equations of motion of the electron as well as the stationary electric field inside the PMT. The geometry of the PMT is imported into the simulation using a CAD file. This file was created with the help of a CT (computed tomography) scan and the geometry is believed to be accurate to the order of $\SI{10}{\micro \meter}$. Note that such simulations have been used previously in \cite{Barrow2017} to investigate the timing properties of afterpulsing of a PMT, though, with a significantly less detailed PMT model. 
In the simulation, the primary electron is defined by the following initial parameters:
\begin{wrapfigure}{b}{0.5\textwidth}
   % \vspace{-1.2cm}
    \centering
    \includegraphics[trim={1.cm 0.24cm 1.075cm 0.4cm},clip,width=0.4\textwidth]{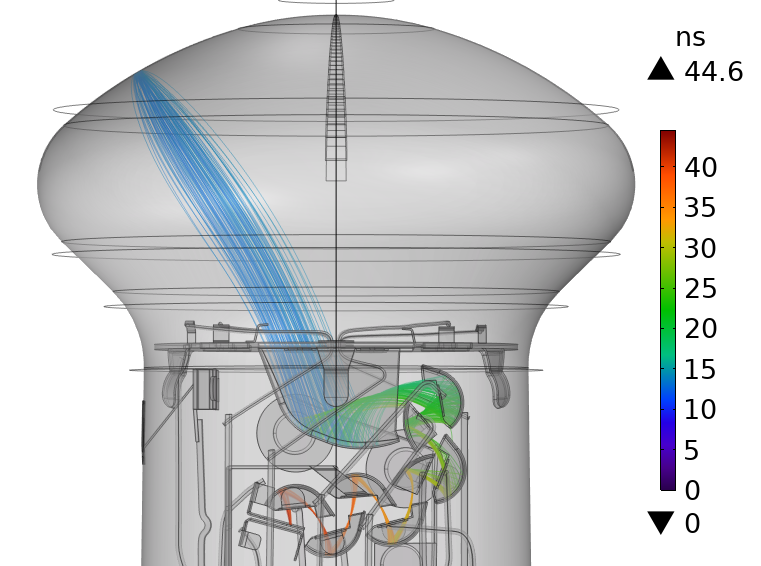}
    \caption{Visualization of 100 simulated electrons, all starting at the same location on the photocathode but with different starting directions.}
    \vspace{-1.2cm}
    \label{fig:Comsol}
\end{wrapfigure}
\begin{itemize}
    \item Position: This can be evenly distributed across the photocathode or set at specific points to simulate scans.
    \item Energy: The energy is sampled from an asymmetric Gaussian distribution. The shape of the distribution is dependent on the wavelength of the incoming photon \cite{AGWeinheimerDoctorialThesis}.
    \item Starting direction: The starting direction is sampled according to Lambert's cosine law \cite{Wright2017}.  
\end{itemize}
For the simulation of a scan measurement, $\SI{10000}{}$ electrons were simulated at different positions on the photocathode. When a primary electron hits a dynode, it produces a single secondary electron at rest (no multiplication is simulated). An example of 100 simulated electrons is shown in figure \ref{fig:Comsol}. The figure shows the simulated electron paths, all starting from the same point at the same time. The color of the electron trajectories describes the time. 
  
\subsection{Results}
\begin{figure}[b]
\vspace{-10mm}
\captionsetup[subfloat]{labelformat=empty}
     \subfloat[\label{subfig:TT_x}]{%
     \includegraphics[width=0.475\textwidth]{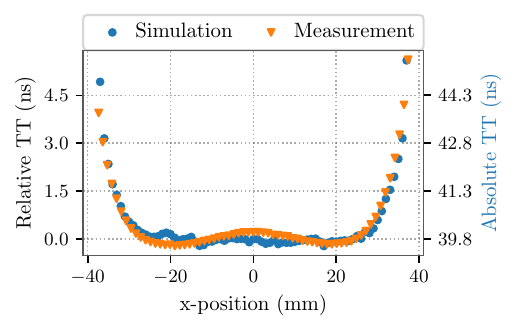}
     }
      \subfloat[\label{subfig:TT_y}]{%
       \includegraphics[width=0.49\textwidth]{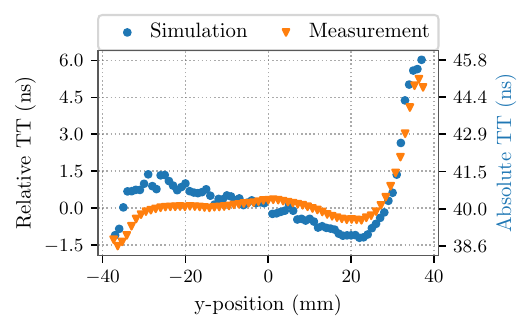}
     }
     \vspace{-0.8cm}
     \caption{Results for the simulation of the scan of the photocathode (blue dots) together with the measured data (orange triangles) reused from~\cite{MartinDoctorialThesis}. The data is normalized to the central region with $r \leq \SI{30}{\milli\meter}$. The left plot shows the scan along the x-direction while the right plot shows the y-direction. The secondary y-axis shows the absolute transit time obtained from the simulation.}
     \label{fig:TT_x_y}
   \end{figure}

The simulation described above is performed along the x and y axis of the PMT photocathode in $\SI{1}{\milli\meter}$ steps to simulate the scan described in Section~\ref{sec:PMT}. The results of this simulation are presented in figure \ref{fig:TT_x_y} (blue dots) compared to the results (orange triangles) of the previously described photocathode scans (see section \ref{sec:PMT}).
 The absolute transit time, shown on the secondary y-axis, is obtained from the simulation and so only valid for the simulated data points. \\
 \\
It can be seen that the simulation generally agrees with the measured transit time depending on the starting position of the photoelectrons, but there are also differences, as in the case of the y-axis between $\SI{-40}{\milli\meter}$ and $\SI{10}{\milli\meter}$. While this matter is currently under investigation, the current hypothesis is that the collection efficiency of the 2nd dynode is lower in this region as electrons can hit the 3rd dynode in the back. This effect is mostly present for electrons starting from this region. This would mean that electrons closer to the second dynode might not get past it. These would be the ones with smaller trajectory length and correspondingly lower transit time. Further explanation could be missing features in the simulation like the proper multiplication within the PMT and that interactions between the electrons are not yet simulated.% These two points can also affect the transit time and lead to differences. Note that while the general behavior is similar the absolute values do not match. Possible reasons for this include different methods of calculating the transit time, as well as the missing features in the simulation. Additionally the simulation only goes from \SI{-38}{\milli\meter} to \SI{38}{\milli\meter}.
\\
\section{Conclusion}\label{sec:Conclusion_Outlook}
Three different methods for a detailed study of PMTs have been presented. The inhomogeneity along the photocathode and its influence on the PMT performance have been studied using PMT surface scans. This led to the idea of two further investigation methods. The first of these delves into the complex refractive index of the photocathode and its thickness, from which the PMT efficiency can be calculated. First ellipsometry measurements yielded promising estimations for the optical parameters of the mDOM PMT. The second method involves simulating the electron trajectories within the PMT and examining how initial conditions namely energy, position, and direction affect PMT performance. This approach has produced results consistent with measurements, thereby simplifying the process of interpreting these observations. Though both studies are still in progress, the preliminary results provide a promising outlook for a more detailed understanding of PMTs.
%\lipsum[3] 

% Bibtex references:
\bibliographystyle{ICRC}
\bibliography{references}

% Alternatively, you can include references by hand:
%\begin{thebibliography}{99}
%\bibitem{...}
%
%\end{thebibliography}

\clearpage

%The following list of authors, affiliations and funding agencies will be updated at the day of submission. The following template is a placeholder generated via https://authorlist.icecube.wisc.edu/icecube on March 25, 2023 and will be updated.
\input{authorlist_IceCube.tex}

\end{document}

%% file: authorlist_IceCube.tex
\section*{Full Author List: IceCube Collaboration}

\scriptsize
\noindent
R. Abbasi$^{17}$,
W. Achtermann$^{43}$,
M. Ackermann$^{63}$,
J. Adams$^{18}$,
S. K. Agarwalla$^{40,\: 64}$,
J. A. Aguilar$^{12}$,
M. Ahlers$^{22}$,
J.M. Alameddine$^{23}$,
N. M. Amin$^{44}$,
K. Andeen$^{42}$
% ...
\\
\\
% ...
$^{12}$ Universit{\'e} Libre de Bruxelles, Science Faculty CP230, B-1050 Brussels, Belgium \\
% ...
$^{17}$ Department of Physics, Loyola University Chicago, Chicago, IL 60660, USA \\
$^{18}$ Dept. of Physics and Astronomy, University of Canterbury, Private Bag 4800, Christchurch, New Zealand \\
% ...
$^{22}$ Niels Bohr Institute, University of Copenhagen, DK-2100 Copenhagen, Denmark \\
$^{23}$ Dept. of Physics, TU Dortmund University, D-44221 Dortmund, Germany \\
% ...
$^{40}$ Dept. of Physics and Wisconsin IceCube Particle Astrophysics Center, University of Wisconsin{\textendash}Madison, Madison, WI 53706, USA \\
% ...
$^{42}$ Department of Physics, Marquette University, Milwaukee, WI, 53201, USA \\
$^{43}$ Institut f{\"u}r Kernphysik, Westf{\"a}lische Wilhelms-Universit{\"a}t M{\"u}nster, D-48149 M{\"u}nster, Germany \\
$^{44}$ Bartol Research Institute and Dept. of Physics and Astronomy, University of Delaware, Newark, DE 19716, USA \\
% ...
$^{63}$ Deutsches Elektronen-Synchrotron DESY, Platanenallee 6, 15738 Zeuthen, Germany  \\
$^{64}$ Institute of Physics, Sachivalaya Marg, Sainik School Post, Bhubaneswar 751005, India \\
% ...

\subsection*{Acknowledgements}

\noindent
The authors gratefully acknowledge the support from the following agencies and institutions:
USA {\textendash} U.S. National Science Foundation-Office of Polar Programs,
U.S. National Science Foundation-Physics Division,
U.S. National Science Foundation-EPSCoR,
Wisconsin Alumni Research Foundation,
Center for High Throughput Computing (CHTC) at the University of Wisconsin{\textendash}Madison,
Open Science Grid (OSG),
Advanced Cyberinfrastructure Coordination Ecosystem: Services {\&} Support (ACCESS),
Frontera computing project at the Texas Advanced Computing Center,
U.S. Department of Energy-National Energy Research Scientific Computing Center,
Particle astrophysics research computing center at the University of Maryland,
Institute for Cyber-Enabled Research at Michigan State University,
and Astroparticle physics computational facility at Marquette University;
Belgium {\textendash} Funds for Scientific Research (FRS-FNRS and FWO),
FWO Odysseus and Big Science programmes,
and Belgian Federal Science Policy Office (Belspo);
Germany {\textendash} Bundesministerium f{\"u}r Bildung und Forschung (BMBF),
Deutsche Forschungsgemeinschaft (DFG),
Helmholtz Alliance for Astroparticle Physics (HAP),
Initiative and Networking Fund of the Helmholtz Association,
Deutsches Elektronen Synchrotron (DESY),
and High Performance Computing cluster of the RWTH Aachen;
Sweden {\textendash} Swedish Research Council,
Swedish Polar Research Secretariat,
Swedish National Infrastructure for Computing (SNIC),
and Knut and Alice Wallenberg Foundation;
European Union {\textendash} EGI Advanced Computing for research;
Australia {\textendash} Australian Research Council;
Canada {\textendash} Natural Sciences and Engineering Research Council of Canada,
Calcul Qu{\'e}bec, Compute Ontario, Canada Foundation for Innovation, WestGrid, and Compute Canada;
Denmark {\textendash} Villum Fonden, Carlsberg Foundation, and European Commission;
New Zealand {\textendash} Marsden Fund;
Japan {\textendash} Japan Society for Promotion of Science (JSPS)
and Institute for Global Prominent Research (IGPR) of Chiba University;
Korea {\textendash} National Research Foundation of Korea (NRF);
Switzerland {\textendash} Swiss National Science Foundation (SNSF);
United Kingdom {\textendash} Department of Physics, University of Oxford.